\documentclass[prb,twocolumn,showpacs,amsfonts,amssymb,floats,superscriptaddress,aps]{revtex4}
\usepackage[final,dvips]{epsfig}
\usepackage{graphicx} 

\newcommand{\pd}{{\phantom{\dagger}}}
\def\ket#1{|#1\rangle}

\begin{document}
\title[]{Multiple-charge transfer and trapping in DNA dimers}
\author{
Sabine Tornow}
\affiliation{\mbox{Institut f\"ur Mathematische Physik, TU Braunschweig, 38106 Braunschweig, Germany}}
\author{
Ralf Bulla}
\affiliation{\mbox{
Institut f\"ur Theoretische Physik, Universit\"at zu K\"oln, 50937 K\"oln, Germany}}
\author{
Frithjof B. Anders}
\affiliation{\mbox{ 
Theoretische Physik II, TU Dortmund,
44221 Dortmund}}
\author{
Gertrud Zwicknagl}
\affiliation{\mbox{Institut f\"ur Mathematische Physik, TU Braunschweig, 38106 Braunschweig, Germany}}
\date{\today}

\begin{abstract}

We investigate the charge transfer characteristics of one and two excess charges in a DNA base-pair dimer
using a model Hamiltonian approach. 
The electron part comprises diagonal and off-diagonal Coulomb matrix elements
such a correlated hopping and the bond-bond interaction, which were recently 
calculated by Starikov [E. B. Starikov, Phil. Mag. Lett. {\bf 83}, 699 (2003)] for different DNA dimers. 
The electronic degrees of freedom are 
coupled to an ohmic or a super-ohmic bath serving as dissipative environment. 
We employ the numerical renormalization group method in the nuclear tunneling regime 
and compare the results  to Marcus theory for the thermal activation regime.
For realistic parameters, the rate that at least one charge is transferred from the donor to the acceptor  in the subspace of two excess electrons significantly exceeds the rate in the single charge sector. Moreover, the dynamics
is strongly influenced by the Coulomb matrix elements. We find sequential and pair transfer as well
as a regime where both charges remain self-trapped. The transfer rate reaches its maximum
when the difference of the on-site and inter-site Coulomb matrix element is equal to the reorganization energy
which is the case in a GC-GC dimer. Charge transfer is completely suppressed 
for two excess electrons in AT-AT in an ohmic bath and replaced by damped 
coherent electron-pair oscillations in a super-ohmic bath. 
A finite bond-bond interaction $W$ alters the transfer rate: it increases as function of $W$ when the effective
Coulomb repulsion exceeds the  reorganization energy (inverted regime) and decreases for smaller 
Coulomb repulsion.

\end{abstract}
\maketitle
\section{Introduction}
Understanding charge transfer dynamics in DNA is of fundamental importance for biochemistry
and molecular electronics. \cite{Chakra,Schuster,Wagenknecht,Starikov2,Endres:2004vg} 
It can generate chemical reactions over
long distances and has biological relevance in the formation of
oxidative damage.
Charge transfer  in DNA also plays an important role in
anti-cancer drugs,\cite{platin} electrochemical
readout of micro arrays and molecular-electronic devices. 
Whereas DNA conductivity is still disputed,
there is no doubt that electron transfer through DNA double strands takes place. \cite{Giese:2002le,Delaney:2003wo}
It is generally believed that the electron or hole
migrates by super-exchange tunneling or
multistep hopping in a donor-bridge-acceptor system depending
on the energy difference between the bridge and donor/acceptor \cite{Wagenknecht}  
and the fluctuation of the environment. \cite{gutierrez:208102,Bulla:2006tx}

Most of the current understanding of electron transfer is derived from a single-excess charge
where only two states are needed to model  a donor-acceptor system. 
At high temperatures the transfer is governed by  thermal activation. 
The rate can be
described in the non-adiabatic limit by the semi-classical Marcus
theory \cite{marcus:966}  and  is determined by the hopping
integral $\Delta$, the energy difference between product and reactant states $\epsilon$, the reorganization
energy $E_{\alpha}$ and the temperature $T$.

At low temperatures, however, the transfer rate does not vanish but rather remains finite and temperature independent:  the  transfer is governed by quantum fluctuations and involves nuclear tunneling.\cite{DeVault:1984rg}
In this regime the transfer is modeled by an electronic two-level system\cite{gutierrez:208102,Barone:1992sw}  
coupled to a dissipative environment  as in the spin-boson model\cite{Leggett:1987gc} which has to be treated quantum mechanically. One can identify three regions of dynamics depending on the coupling 
to the bosonic bath: coherent (damped oscillations), incoherent (exponential relaxation) transfer and self-trapping.

However, in many cases charge transfer involves multiple excess charges which are correlated. \cite{Evans:2008ij,Muhlbacher:2005gn,tornow:035434}
Their transfer characteristics is changed drastically in comparison to the case of a single excess charge
and requires a description by more than two states.\cite{zusman:165,Zusman:1997qe,Petrov:2003qt}
Two charges can migrate as pairs, sequentially or can be self-trapped. 
Their dynamics are governed not only by the coupling to a dissipative bath but also 
by the local and non-local Coulomb interactions ($U$ and $V$, respectively.)\cite{refId,Muhlbacher:2005gn,tornow:035434} 

In this paper
we study the charge transfer dynamics of two electrons or two holes in different DNA 
base-pair dimers based on  Guanine/Cytosine (GC)
or Adenine/Thymine (AT) pairs. We have used  the base-pair specific Coulomb matrix elements as 
calculated with ab-initio methods.\cite{Starikov}  We assume that 
two charges are initially prepared on one base-pair monomer (donor) -- this could be achieved, e.g., by voltage gating. Then, the transfer of those two charges occurs either in a concerted manner, 
where an electron pair or hole pair is transferred between the donor (D) and the acceptor (A), 
or both charges migrate sequentially but correlated. 
This transfer of the first electron facilitates or hinders the transfer of the second one due to the electronic correlations.  
We find that depending on the Coulomb matrix elements  a second charge can lead
even to self-trapping or, in the other extreme, to partially activationless transfer.

Recently, conductance measurements of single DNA molecules were performed in aqueous solution. \cite{Xu:2004wq} The solution preserves the native confirmation and the transport properties can be associated with charge transfer reactions at least for weak coupling to the leads.\cite{Nitzan:2001ws} 
At voltages equal to the on-site Coulomb interaction we expect that more than one excess charge 
is relevant for the charge transport through a molecular device. Consequently, the 
transport characteristics is altered considerably. \cite{TornowGoldenrule} 
Furthermore, two-electron transfer in Guanine due to strong oxidants has been discussed in connections to in vivo lesions. \cite{GeneviEve-Pratviel:2006kn}

Our detailed study of the base-pair specific transfer characteristics of multiple excess charges 
for realistic systems reveals the connection between the microscopic parameters and the transfer efficiency.
Previously,  only an electronic  model with density-density interactions coupled 
to an ohmic bath has been investigated.\cite{tornow:035434}
In  this paper we extend the calculations to a more general electronic model which
includes  an additional bond-bond interaction (pair-transfer) $W$ previously neglected
as well as a bond-charge interaction $X$
(correlated hopping.) These matrix elements have been  calculated for different DNA base-pairs\cite{Starikov} 
using ab-initio methods and serve as input for our calculations.

We have found that the additional bond-bond interaction $W$ favors
concerted electron transfer from the donor to the acceptor but can
also lead to a change of the ground state of the dimer. 
The parameter $W$ has a profound impact 
on the ground state of the electronic subsystem with two excess charges: it lifts
the degeneracy of singlet and the triplet state. The latter becomes the ground state for $W>0$ and a positive effective
Coulomb matrix element. Since the Hamiltonian considered here is spin-conserving, the system can only decay into an
metastable steady state in this regime  starting with two excess charges on the donor in a spin singlet state.

Another important application of charge transfer are DNA biosensors. During the sensing process, 
electron transfer is used to produce electronic signals. \cite{Fan2005186}
In general the signal magnitude depends on the efficiency of the electron transfer through the DNA as well as the number of electrons. It has been suggested that the transfer of multiple electrons would  enhance
those signals.\cite{Fan2005186}
However, our work shows that the transfer rate for a single electron in systems with two excess charges depends on the environment as well as the Coulomb matrix elements: it may be much larger 
or much smaller than the transfer rate of a single excess charge depending on the microscopic parameters.

For temperatures  lower than the activation energy the bath must be treated quantum-mechanically
to  calculate the time-dependent population probabilities and the transfer rates.
In this regime, we employ the time-dependent numerical renormalization group (TD-NRG) approach\cite{anders:245113,Anders:2005lm,Krishna-murthy:1980bz,tornow:035434,bulla:045122,Bulla:2003pw,bulla:395}
whereas in the thermal activation regime we compute the transfer rates using
the Marcus theory.\cite{marcus:966,Nitzan_book}

The paper is organized as follows. In section \ref{model} we
introduce the microscopic model as well as both methods  for high and
low temperature regimes. We present a detailed description of the
single and multiple charge dynamics in Sec. \ref{singlecharge}
and Sec. \ref{multcharge}, respectively.  A summary and conclusions of our
results is given in Sec.~\ref{sum}.

\section{Model and methods}

\subsection{The model}

\label{model}

\begin{figure}
\vspace*{-2cm} \hspace*{0cm} \epsfxsize=18cm
\centerline{\epsffile{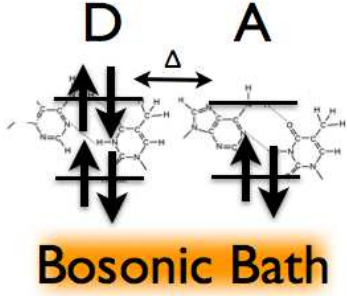}}
\vspace*{-7cm}
\caption{(Color online) Schematic view of the DNA dimer. Each base pair is approximated by a single site with 
a HOMO and LUMO. The electronic degrees of freedom are coupled to the environment comprising internal vibrations and the solvent dynamics. The transfer characteristics of multiple excess charges strongly depends on the base-pair specific Coulomb interactions.} \label{fig_schema_DNA}
\end{figure}

We consider a base-pair dimer represented by a two-site system as donor (D) and acceptor (A) with a LUMO (lowest unoccupied molecular orbital) and a HOMO (highest occupied molecular orbital) coupled to an aqueous environment (bosonic bath) as sketched in Fig.~\ref{fig_schema_DNA}.  
The Hamiltonian of the coupled system consists of three parts
\begin{eqnarray}
H=H_{el}+H_b+H_{el-b} \ , \label{eqn_gen}
\end{eqnarray}
where $H_{el}$ denotes the electronic part of the Hamiltonian, $H_b$ the bath and $H_{el-b}$ the coupling between both subsystems.
Each base pair monomer is modelled as a single site with one orbital (either LUMO or HOMO depending on excess holes or electrons). The energy gap between HOMO and LUMO is about $4 eV$. \cite{Endres:2004vg} 
$H_{el}$ has been derived in  Ref. [\onlinecite{Starikov}]  and corresponds to 
a general two-site extended Hubbard model:
\begin{eqnarray}
H_{el}=\sum_{i=D,A;\sigma} \epsilon_i c_{i \sigma}^{\dag} c_{i \sigma}
+\sum_{\stackrel{\sigma,i,j}{ ( i \neq j)}} \Delta_{ij} c_{i
\sigma}^{\dag} c_{j \sigma} \\ \nonumber +
\sum_{i,j,k,l,\sigma,\sigma'} \left \langle
i j \left | \frac{e^2}{r} \right |kl \right \rangle  c_{i \sigma}^{\dag} c_{j \sigma'}^{\dag}
c_{l \sigma'} c_{k \sigma} \ \ ,
\label{eq:Hel}
\end{eqnarray}
where $c_{i\sigma}$ and $c^{\dagger}_{i\sigma}$ denote annihilation
and creation operators for fermions with spin $\sigma$ in an orbital
localized at site $i$. 
The first and second term are the on-site energy and the transfer
integral or single-charge hopping for nearest neighbors,
respectively. The latter defines the kinetic energy. 
The matrix elements   $\left \langle
i j \left | \frac{e^2}{r} \right |kl \right \rangle$  account for the 
Coulomb repulsion between two electrons. The dominant
interactions in the DNA dimers\cite{Starikov} are
\begin{eqnarray}
\label{Coul}
& & U_i=\left \langle
i i \left | \frac{e^2}{r} \right |ii \right \rangle \\ \nonumber
& & V_{ij}=\left \langle
i j \left | \frac{e^2}{r} \right |ij \right \rangle \\ \nonumber
& & X_{ij}= \left \langle
i i \left | \frac{e^2}{r} \right |ij \right \rangle \\ \nonumber
& & W_{ij}= \left \langle
i i \left | \frac{e^2}{r} \right |jj \right \rangle \ \ .
\end{eqnarray}
The first two matrix elements denote the on-site and nearest
neighbor electron-electron interaction, respectively. The parameter $X$
labels the bond-charge interaction and is often called
correlated hopping since it depends on the density of the nearest
neighbor. The last term is the bond-bond interaction which induces the pair hopping 
and the exchange interaction.

For the donor-acceptor pair, we arrive at the following Hamiltonian:
\begin{eqnarray}
H_{el}&=&\sum_{\sigma; i=D,A} \epsilon_i n_{i \sigma} -\Delta
\sum_{\sigma} \left( c_{D \sigma}^{\dag} c_{A \sigma}+c_{A \sigma}^{\dag} c_{D \sigma} \right) \nonumber \\
&+& U \sum_{i =D/A,\sigma} n_{i \sigma } n_{i -\sigma}  + V
\sum_{\sigma,\sigma'} n_{D \sigma} n_{A \sigma'} \nonumber \\
&+&X \sum_{\sigma}\left( c_{D \sigma}^{\dag} c_{A \sigma} + c_{A
\sigma}^{\dag} c_{D \sigma} \right) \left(n_{D -\sigma}+n_{A
-\sigma} \right) \nonumber \\
&+& W\sum_{\sigma}  \left( c_{D \sigma}^{\dag} c_{A
-\sigma}^{\dag} c_{D -\sigma} c_{A \sigma} + c_{A \sigma}^{\dag}
c_{D -\sigma}^{\dag} c_{A -\sigma} c_{D \sigma} \right.  \nonumber \\
&+& \left.  c_{D \sigma}^{\dag} c_{D
-\sigma}^{\dag} c_{A -\sigma} c_{A \sigma} + c_{A \sigma}^{\dag}
c_{A -\sigma}^{\dag} c_{D -\sigma} c_{D \sigma} \right) 
\label{eq:model}
\end{eqnarray}
with the number operator $n_{i, \sigma}=c_{i,\sigma}^{\dag} c^\pd_{i,\sigma}$. The last two terms correspond to hopping of electron pairs. In our previous study,\cite{tornow:035434}
only the matrix elements $U$ and $V$ have been taken into account.
Here, we will investigate (i) the influence of the additional matrix elements $X$ and $W$ onto the transfer
dynamics using (ii) realistic values for $U,V,X,W$ as well as for the coupling to the environment as will be 
discussed in the following.

The bosonic bath comprises the internal vibrations and solvent dynamics and leads to dephasing and 
dissipation of the electronic system. 
The Hamiltonian $H_b$
\begin{eqnarray}
H_b=\sum_{n} \omega_{n} b_{n}^{\dagger} b^\pd_{n} 
\end{eqnarray}
models the free bosonic bath.
The displacements of the  polaronic bath  couple to the change of the electron density,  
\cite{Leggett:1987gc,garg:4491}
\begin{eqnarray}
H_{el-b}=\sum_{\sigma}\left(n_{D,\sigma}-n_{A,\sigma}\right) \sum_{n}
        \frac{\lambda_n}{2} \left(  b_{n}^{\dagger} + b^\pd_{n}
        \right),
\end{eqnarray}
and
give rise to an attractive contribution to the on-site Coulomb interactions. The renormalized interaction is defined by
\begin{eqnarray}
\tilde{U}_{\rm eff}= U-V-\sum_{n}\frac{\lambda_n^2}{\omega_n} ,
\end{eqnarray}
which determines the energy difference between the ionic and covalent states for $W=0$
and includes the influence of the environment.

\begin{figure}
\vspace*{0cm} \hspace*{0.0cm} \epsfxsize=10cm
\centerline{\epsffile{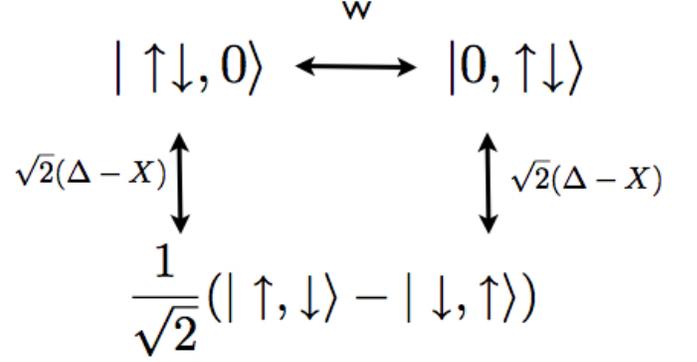}}
\vspace*{-1cm}
\caption{Schematic view of the relevant states: The ionic states $|\uparrow\downarrow,0 \rangle$ and $|0,\downarrow \uparrow \rangle$ as well as the covalent singlet state $\Psi_s$. They are connected by the renormalized hopping $\Delta-X$ as well as pair hopping $W$. 
$\ket{\Psi_t}$ is not connected to the other states. The energy difference between the ionic states and 
$\ket{\Psi_s}$ is $\tilde{U}_{\rm eff}-W$. Only the ionic states are coupled to the bosonic bath.} \label{fig_schema_states}
\end{figure}

The Hamiltonian of the isolated dimer conserves the total charge
$Q=4-n$, 
where $n$ is the total number of electrons. We consider the 
case $Q=2$ (two holes (2h) in the HOMO), $Q=-2$ 
(two electrons (2e) in the LUMO) as well as $Q=1$ and $Q=-1$ (single hole in the HOMO and single electron in the LUMO, respectively). 

Furthermore,  the square of the total spin as well as its $z$-component $S_z$ are conserved by the 
Hamiltonian Eq.~(\ref{eq:model}). The subspace of two holes/electrons ($|Q|=2$) and $S_z=0$ (for $|S_z|=1$ no transfer is
possible) is spanned by 4 states, two ionic states with two electrons/holes
on the donor $| \uparrow \downarrow, 0 \rangle $ or on the acceptor $| 0, \uparrow \downarrow \rangle $ and the two
covalent states 
\begin{eqnarray}
\ket{\Psi_s} &=&\frac{1}{\sqrt{2}}(|\uparrow,\downarrow \rangle-|\downarrow,\uparrow \rangle) \nonumber \\
\ket{\Psi_t} &=&\frac{1}{\sqrt{2}}(|\uparrow,\downarrow \rangle+|\downarrow,\uparrow \rangle) \ . 
\end{eqnarray}
The latter states correspond to the singlet and triplet states, respectively.
$\ket{\Psi_t}$ is an eigenstate of $H_{\rm el}$ with energy $E_t=-W$ and is decoupled from the other three states displayed in Fig.~\ref{fig_schema_states}. It is the ground state for $4W>\sqrt{U^2+16 (\Delta-X)^2}-U$ ($\alpha=0$). 

In the subspace of $|Q|=2,S_z=0$,  the total number of electrons per spin, $n_{D \sigma}+n_{A\sigma}$, is constant, 
and the correlated hopping amplitude $X$ renormalizes the single particle hopping to $\Delta \rightarrow \Delta -X$.

For one electron or hole ($|Q|=1$) and $|S_z|=1/2$ the model is equivalent to the spin-boson model.
The charge is able to migrate
between the two states $| \bullet  ,0 \rangle$ and $| 0, \bullet \rangle$ with one electron/hole on the donor and acceptor, respectively.

The coupling between electronic subsystem and bosonic bath is specified by the bath
spectral function\cite{Leggett:1987gc,Weiss_book}
\begin{equation}
  J(\omega) = \pi   \sum_{n}
\lambda_{n}^{2} \delta\left( \omega -\omega_{n} \right) \ .
\end{equation}
$J(\omega)$ characterizes the bath and the 
system-bath coupling and reflects intramolecular vibrations and the solvent (water). 
The spectral function can be calculated by molecular dynamics simulations\cite{vladimirov:194515} for a two-sphere DA model embedded in an aqueous medium and shows a power law behavior at low frequencies. 
The asymptotic low-temperature behavior is determined by the low energy
part of the spectrum. Discarding high-energy details,
we use the standard parametrization\cite{Leggett:1987gc,Weiss_book} 
\begin{eqnarray}
J(\omega) = \left \{   \begin{array}{cc} 2 \pi \alpha
\omega^s \omega_c^{1-s}  &  0 < \omega < \omega_c \ ,
\\ 0 & {\rm otherwise} \ , \end{array}  \right.
\end{eqnarray}
where $\alpha$ is a dimensionless constant which characterizes the dissipation strength and $\omega_c$ is the cutoff frequency of the spectral function.

We investigate an ohmic bath ($s=1$) and $s=2$ as an example for the super-ohmic dissipation.\cite{Kleink} 
As an realistic estimate for the cut-off frequency $\omega_c$ we use $0.5 eV$ which correspond to the frequency
of the OH stretch vibrations at about $3700 cm^{-1}$.\cite{Praprotnik:2004yf}

The classical reorganization
energy $E_{\alpha}$ for the single-electron transfer $|\bullet,0 \rangle \rightarrow |0, \bullet \rangle $,
which measures the energy relaxation following a sudden electronic transition,
\begin{equation}
  E_{\alpha} =   \sum_n \frac{\lambda_n^2}{ \omega_n}=
\frac{1}{\pi}\int_0^\infty d\omega\, \frac{J(\omega)}{\omega} =
  \frac{2\alpha \omega_c}{s} \ ,
\label{reo}
\end{equation}
has been determined\cite{Wagenknecht} for DNA in water as approximately $E_{\alpha}=0.5eV$. 
Therefore, we have used $\alpha=0.5$ and $\alpha=1$ for the ohmic bath and super-ohmic 
($s=2$) bath, respectively, for modeling the coupling between the DNA dimers and the environment.
The reorganization energy for the transfer $|\!\uparrow \downarrow,0 \rangle \rightarrow \ket{\Psi_s} $ 
in the sector of two excess electrons is given by $E_{ \alpha 1}=4 E_{\alpha }=2eV$ and for the transfer $|\! \uparrow \downarrow,0 \rangle \rightarrow | 0,\uparrow \downarrow \rangle $ given by  $E_{ \alpha 2}=4 E_{\alpha 1}=8eV$. 

The single-charge hopping $\Delta$ as well as pair-hopping $W$ are renormalized by the bosonic bath.  Because only the ionic states are coupled to the bath, the pair-hopping is much stronger reduced than $\Delta$ which connects ionic and covalent states. 
The exchange interaction, however, which is also proportional to $W$ remains unaltered by the bath 
since it does not change the occupancy of the orbitals.

\begin{figure}
\vspace*{0cm} \hspace*{0cm} \epsfxsize=9cm
\centerline{\epsffile{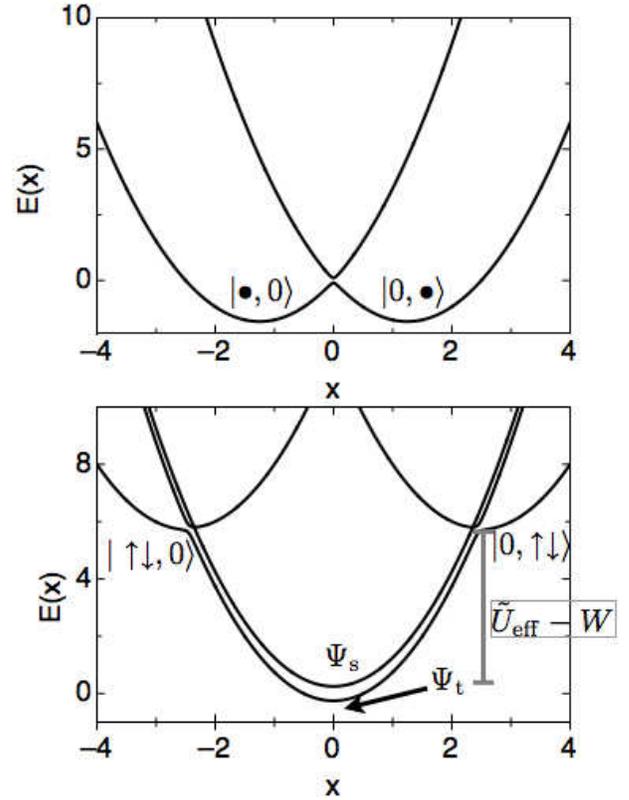}}
\caption{Schematic potential surfaces $E(x)$ for the lowest states of the model in the one-electron
(upper panel) and two-electron subspace (lower panel) for the displacement $x$ in arbitrary units. The energy difference between the ionic states ($| \uparrow \downarrow,0 \rangle $, $|0, \uparrow \downarrow \rangle $)  and covalent singlet state $\Psi_s$ corresponds to the difference between the renormalized Coulomb interaction $\tilde{U}_{\rm eff}$ and bond-bond interaction $W$.}
\label{fig_schema}
\end{figure}

In the following sections, the four time-dependent population probabilities,
defined as the expectation values of the operators
\begin{eqnarray}
n_D &=& | \bullet,0 \rangle  \langle  0,\bullet | \nonumber \\
d_D &=&|\uparrow \downarrow, 0 \rangle \langle 0, \uparrow \downarrow| \nonumber \\
d_A &=&|0,\uparrow \downarrow \rangle \langle \uparrow \downarrow, 0| \nonumber \\
n_{DA}&=&| \uparrow, \downarrow \rangle \langle  \downarrow, \uparrow| + |\downarrow, \uparrow \rangle \langle\uparrow,\downarrow \!| ,
\end{eqnarray}
are calculated. They measure the single and double occupancy on the donor and the acceptor site (ionic states)
as well as the equally populated states (covalent states).

\subsection{The Marcus theory}

The Marcus theory of charge transfer is connected to the quantum mechanical treatment
of the bosonic bath in Hamiltonian (\ref{eqn_gen})
by a mean-field decoupling of the  bosonic bath modes 
from the electronic degrees of freedom. 
The local displacement $\hat x = \sum_n \lambda_n(b^\dagger_n + b_n)$ is replaced by its expectation value $x$, which is also called the reaction coordinate. The resulting
potential surfaces $E(x)$ for  the two states $D^-A=| \bullet,0 \rangle   $ and  $DA^-=| 0,\bullet\rangle $  
in the $|Q|=1$ sector are plotted schematically in the upper panel of Fig.~\ref{fig_schema}. The   
potential surfaces $E(x)$ of the four states  in the $|Q|=2$ sector,
$D^{2-}A=| \uparrow \downarrow ,0 \rangle $,  $DA^{2-}=|0, \uparrow \downarrow  \rangle $ and $ D^{-}A^-= \ket{\Psi_s}$ or $ \ket{\Psi_t} $ are depicted  in the lower panel of Fig.~\ref{fig_schema}.

The single-electron transfer from $D^-A$ to $DA^-$ is determined by the hopping $\Delta$ and the 
coupling constant $\alpha$ (or the reorganization energy $E_{\alpha}$) .
For $|Q|=2$, the transfer additionally depends on 
$\Delta-X$ and $W$. Moreover, the effective bath coupling constant is four times larger due to a factor 2 in the coupling term. As a consequence one would expect a slower transfer compared to  the $|Q|=1$ case. 
However, the transfer of two charges is strongly affected by the energy difference between the ionic and covalent singlet states $\tilde{U}_{\rm eff}-W$. The transfer of one charge can be even activationless if the difference $\tilde{U}_{\rm eff}-W$ is equal to the reorganization energy $E_{\alpha 1}$ as indicated in the lower panel of Fig.~\ref{fig_schema}
at finite values of $x$. 

To address the thermal activation regime where the temperatures are larger than the activation energy ($T>\frac{\left(E_{\alpha1} - \tilde{U}_{\rm eff}-W\right)^2}{4 E_{\alpha1}}$),  the dissipative environment is  treated classically, 
and the rates are obtained from  Fermi's
golden rule. Generally the Marcus rate is given by
\begin{eqnarray}
k_{i \rightarrow f}=\frac{2 \pi}{\hbar} |\langle f| H_{DA} | i \rangle|^2 F(\Delta E_{if}),
\label{fgr}
\end{eqnarray}
where in the Franck-Condon integral
\begin{eqnarray}
F(\Delta E_{if})=\frac{1}{\sqrt{4 \pi E_{\alpha}^{i \leftrightarrow
f} k_B T}} e^{- \frac{\left(\Delta E_{if}-E_{\alpha}^{i
\leftrightarrow f}\right)^2}{4 E_{\alpha}^{i \leftrightarrow f} k_B
T}}
\end{eqnarray}
the energy difference between state $i$ and $f$, $\Delta E_{if}$,  
and  the corresponding reorganization energy 
$E_{\alpha}^{i \leftrightarrow f}$ -- as described   after Eq.~(\ref{reo}) --  enters. 
The population probabilities  are determined by kinetic equations
\begin{eqnarray}
\dot{P}_s=\sum_{r} k_{r \rightarrow s} P_r-\sum_{r} k_{s \rightarrow r} P_s,
\label{kinetic}
\end{eqnarray}
where $P_s$ is the probability that the system is found in state $s$ and
$k_{r \rightarrow s}$ is the Marcus
rate of the transition from state $r$ to state $s$.

\subsection{Time-dependent numerical renormalization group approach}

At low temperatures  ($T\ll\frac{\left(E_{\alpha1}-\tilde{U}_{\rm eff}-W\right)^2}{4 E_{\alpha1}}$) 
the bosonic bath must be treated  quantum mechanically. In this
regime, 
coherent quantum oscillations 
may be present, and we  have calculated the 
time-dependent population probabilities using the TD-NRG. \cite{anders:245113,Anders:2005lm,tornow:035434} 

For the models studied in this work, the impurity couples to a bosonic
bath, for which the standard NRG approach has to be modified as described in
Refs.~\onlinecite{bulla:045122,Bulla:2003pw,bulla:395}.
The basic features of the bosonic NRG are 
as follows: 
(i) the logarithmic discretization of the bath spectral function $J(\omega)$ in intervals $[\Lambda^{-(n+1)},\Lambda^{-n}]$, with $n = 0, 1, . . . , \infty$ 
and $\Lambda > 1$ the NRG discretization parameter (all the 
results shown in this paper are calculated using $\Lambda = 2$); within each of these intervals only 
one bosonic degree of freedom is retained as a representative of the continuous set of degrees 
of freedom. 
(ii) The mapping of the resulting Hamiltonian onto a semi-infinite chain. 
(iii) The 
iterative diagonalization of the chain-Hamiltonian via successively adding one site to the chain.

For the results shown here, we use $N_b=8$ bosonic basis states for each chain link
and retain the $N_s=100$ lowest eigenstates after each
iteration.  
The set of all 
discarded states eliminated during the NRG procedure form 
a complete basis of the semi-infinite chain, which is also an 
approximate eigenbasis of the Hamiltonian. 
In one NRG run we compute the initial density matrix,\cite{anders:245113,Anders:2005lm,tornow:035434} 
in the second run the approximate eigenbasis of the Hamiltonian Eq.~(\ref{eq:model}).  Since the Hamiltonian  usually does not commute with the initial density matrix, the time
evolution of the density matrix governs the time dependence of all expectation values.

The real-time dynamics of the population probabilities has been calculated by summing over
all contributions  from the sequence of reduced density matrices 
which contain the information of decoherence and dissipation at a given energy scale $\omega_c\Lambda^{-n}$.
To simulate a bath continuum we have averaged over different bath discretizations (we used eight different realizations).
For more details on the approach see Refs.~[\onlinecite{anders:245113,Anders:2005lm}].

\section{Subspace of a single charge}
\label{singlecharge}

Before analyzing the transfer of multiple excess
charges we briefly review  the single-charge transfer. The two-state
system is prepared such that the donor is occupied by one
electron/hole at $t=0$. The transfer of a single charge is not effected by the Coulomb matrix elements and Eq.~(\ref{eq:Hel}) reduces to the first two terms which is equivalent to the spin-boson model. 
We use a typical hopping matrix element$\Delta=0.4 \omega_c=0.2 eV$ as obtained by ab-inito methods\cite{Starikov,Chakra}
and  $\alpha=0.5$  as estimated above.

\begin{figure}[tb]

\includegraphics[width=0.48\textwidth]{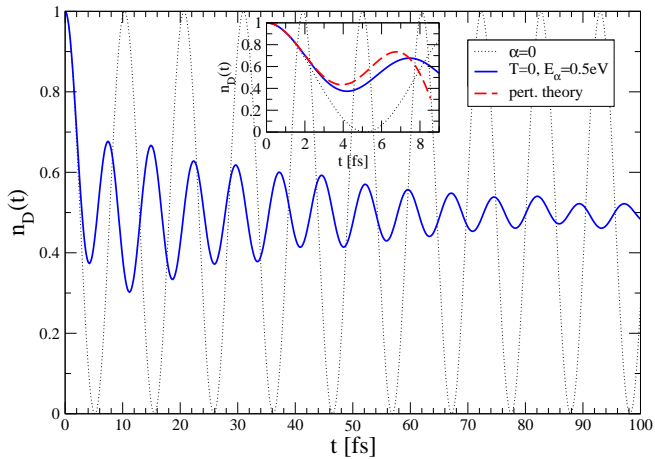}

\caption{(color online) Time-dependent population probability of the donor site, $n_D(t)$ in the subspace of a single charge. The system
is prepared such that at time t = 0 the charge occupies the donor.
The results are shown for $E_\alpha=0.5eV$ (blue), a decoupled charge (dotted line) for $\alpha=0$.
The inset shows a comparison with simple second-order perturbation
theory in $\alpha$
 which is only valid for very short 
time scales. } \label{fig_single}
\end{figure}

The population of the donor $n_D(t)$  calculated with the TD-NRG  for $T\to 0$
is shown in Fig.~\ref{fig_single}.
After about 200 fs the system relaxes to its ground state $n_D=\frac{1}{2}$. 
The frequency of the oscillations increased due to the strong coupling to the bosonic bath, which is counterintuitive at first.

Usually, a reduction of the frequency  in a damped harmonic oscillator  is expected due to friction. This
has  also been found in the spin-boson model in the limit\cite{Leggett:1987gc}  of $\Delta/\omega_c\to 0$, when the 
bare oscillation frequency is much smaller than the bath cutoff energy.  In that limit, all coherent
oscillations are suppressed once $\alpha\to 0.5$. 

By comparing the result to the dynamics of a decoupled spin  ($\alpha=0$, dotted line) we find   indeed 
a slower spin-decay at ultra-short time scales of $t<4$fs -- see inset of Fig.~\ref{fig_single}. 
However, the realistic  DNA parameters  yield $2\Delta\approx \omega_c$: the bare frequency of oscillation is of the order of the cutoff of the bosonic bath. Now, band edge effects become important, and the strong coupling to the bath ($\alpha=0.5$) lead to a level splitting in the first RG steps (level repulsion).  Some local excitation energies 
are shifted above the cutoff $\omega_c$; other contributions are renormalized further down to low frequencies in each RG step. This is reflected in the Fourier transformation of $P(t)$ -- not shown here -- which  peaks 
at a frequency of approximatly $1.13 \omega_c$; however, it  also comprises a  
second broad contribution 
which has its maximum at a reduced frequency of $\omega\approx 0.18\omega_c$.

The  analytic solution obtained in second-order perturbation theory in the bath-coupling $\lambda_q$ 
agrees very well with the ultra-short dynamics of TD-NRG.
This is depicted in  the inset of Fig.~\ref{fig_single}: the perturbative and the TD-NRG solution are identical for $t<3$fs for ohmic and super-ohmic baths -- not shown here. 
In fact the perturbative solution shows even a stronger renormalization of the frequencies than the TD-NRG. However, the solution becomes less reliable for $t> 6$fs due to the strong coupling effects ignored in the perturbation theory, which is only valid at very short time scales. The origin of this enhancement can be traced to a factor $\Delta/(\omega^2 -\Delta^2)$ in the analytic integral kernel which causes a sign change of the contribution 
for $\omega <\Delta$ to   $\omega > \Delta$. The position of $\Delta$
relative to the cutoff $\omega_c$ and the power $s$ of $J(\omega)$
determines the sign of the frequency renormalization.

The spin-boson model is usually only investigated in the limit $\Delta/\omega_c\to 0$.
The DNA dimer system, however, appears to be in the complementary regime
where  $2\Delta \to \omega_c$. In the NRG as well as in the analytical solution, 
we find coherent oscillations which prevail well above $\alpha=1$ (not shown here).

\begin{table}
\begin{tabular}{c |  c}
\hline  Dimer &  $\tilde{U}_{\rm eff}$   \\ 
\hline  \hline   AT-AT (2h)&  0.1 eV  \\ 
\hline    AT-AT (2e) &   -1.0 eV  \\ 
\hline    GC-GC (2h)&   2.0 eV  \\ 
\hline    GC-GC (2e)&  1.0 eV  \\ 
\hline 
\end{tabular}
\caption{Coulomb interaction parameter taken from Ref.~[\onlinecite{Starikov}]. The other parameters are $\Delta=0.2 eV$, $X=0.1 eV$, $\omega_c=0.5 eV$, $E_{\alpha}=0.5$, $E_{\alpha1}=2 eV$ and $E_{\alpha1}=8 eV$. We vary $W$ between $-0.25 eV$ and $0.25 eV$.  } 
\label{table:I}
\end{table}

\section{Subspace of two charges}
\label{multcharge} 

To investigate the influence of
the diagonal and the additional off-diagonal Coulomb interactions on the transfer
of two charges, we initially  prepare the system such that two electrons/holes are localized on the donor
at time $t=0$. This can  be achieved experimentally, e.g., through
radiolytic or photolytic excitation, strong oxidation or voltage
gating. 

The  hopping matrix element $\Delta$ is renormalized to $\Delta-X$  
in the presence of a finite correlated hopping matrix element $X$ in the subspace of two charges.
The base-pair specific Coulomb interaction used in the following are stated in table I 
and are taken from Ref.~[\onlinecite{Starikov}]. 
Since we are mainly interested in the influence of the Coulomb interactions onto the transfer rates, 
we set the hopping matrix element to a typical value\cite{Starikov} $\Delta=0.2 eV$ and a correlated hopping parameter $X=0.1 eV$ for all base-pairs:\cite{Starikov} the dynamics of a single charge
is governed by the energy scale of $0.1eV$.

It has been argued\cite{Starikov} that  the bond-bond interaction $W$ is either small or negligible. Therefore, 
we have investigated the influence of $W$ onto the electron transfer dynamics by varying $W$ 
between $W=-0.25 eV$ and $W=0.25eV$.  In the next section, we begin with $W=0$.

\subsection{Zero bond-bond interaction}

\begin{figure}[t]

\vspace*{-10mm}

\includegraphics[width=0.52\textwidth]{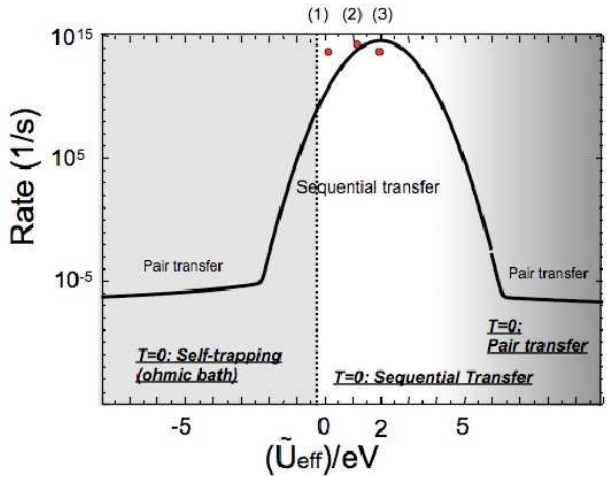}

\caption{(Color online)  Rate $k_{DA}$ as defined in
  Eq.~(\ref{steady}) at high 
temperatures (black line) and low temperatures obtained from NRG (red
points), as a function of $\tilde{U}_{\rm eff}$. The region of sequential and concerted pair transfer is indicated. For $\tilde{U}_{\rm eff}>6 eV$ the pair transfer is followed by a
very slow transfer to the covalent state. At low temperatures three regions are present: Self trapping for an ohmic bath (left of the dotted line), sequential transfer (white region) and pair transfer combined with slow sequential transfer (shaded region for $\tilde{U}_{\rm eff}>0$). The low-temperature rates were obtained by fitting the population probabilities to an exponential function. The points represent the rates for (1) AT-AT(2h), (2) GC-GC (2e) and (3) GC-GC (2h).   
} \label{U4double}
\end{figure}

The Marcus theory is applicable in the temperature regime where the bosonic bath can be replaced by
a single reaction coordinate, and  the physics is governed by thermal activation. The theory
predicts an exponential dependency of the transfer  rates on the energy difference 
$\tilde{U}_{\rm eff}-E_{\alpha 1}$. The concerted transfer of both electrons becomes relevant once the transfer rate between the ionic states exceeds the rate between ionic and covalent states. 
The latter regime, however,  is excluded by the parameters stated in tab.~\ref{table:I} in the DNA dimers considered in
this paper.

The rate $k_{DA}$ that at least one electron is transferred from the donor to the acceptor in the thermal activation regime 
is defined as 
\begin{eqnarray}
k_{DA}=k_{|\uparrow \downarrow, 0 \rangle \rightarrow |0,\uparrow \downarrow \rangle}+k_{|\uparrow \downarrow, 0 \rangle \rightarrow
\ket{\Psi_s}}
\label{steady}
\end{eqnarray}
and is plotted versus $\tilde{U}_{\rm eff}$ as a solid line in Fig.~\ref{U4double}.
The maximum at
$\tilde{U}_{\rm eff}=E_{\alpha 1}$ marks the optimal or activation-less regime and separates 
the normal regime,  $\tilde{U}_{\rm eff}<E_{\alpha 1}$, from the inverted regime, $\tilde{U}_{\rm eff}>E_{\alpha 1}$.
For  $\tilde{U}_{\rm eff}<-2eV$ and $\tilde{U}_{\rm eff}>6eV$ the pair transfer is the main process. 
For $\tilde{U}_{\rm eff}>0$ it is accompanied by a slow sequential transfer to the covalent state.

\begin{figure}[t]

\includegraphics[width=0.5\textwidth]{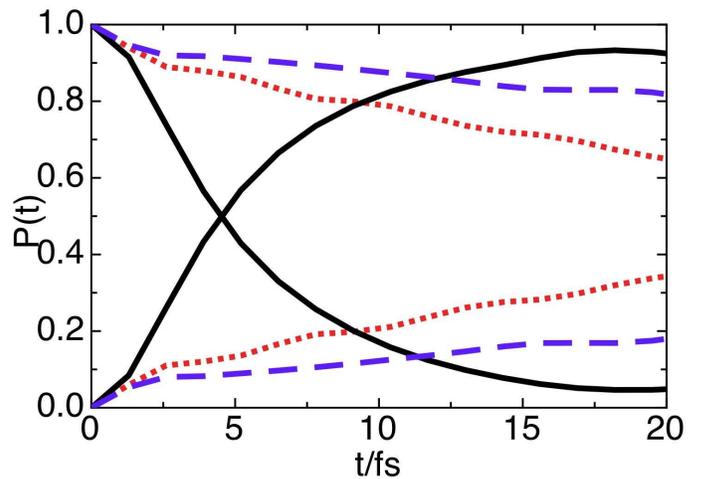}

\caption{(Color online). Low-temperature population probability of the doubly occupied donor $P(t)=d_D(t)$ (starts at one) and the covalent states $P(t)=n_{DA}(t)$ (starts at zero) for AT-AT (2h) (red dotted line), GC-GC (2e) (black solid line) and for GC-GC (2h) (blue dashed line) as a function of time for an ohmic bath ($s=1$) and $W=0$ using the TD-NRG.
At t=0 two electrons are localized at the donor. Two excess electrons on AT-AT 
are self-trapped and not displayed. } \label{fig_QuantumU}
\end{figure}

Using the parameters of tab.~\ref{table:I}, we note that 
the transfer occurs  activationless for two holes in a GC-GC dimer  because $\tilde{U}_{\rm eff}=E_{\alpha 1}$.
The estimate of the transfer rate via Eq.~(\ref{fgr}) at room temperature translates into an ultra-fast 
characteristic time scale of about 10fs. For two electrons on the GC-GC dimer, however, 
the bare $U$ is smaller, and  the condition $\tilde{U}_{\rm eff}<E_{\alpha 1}$ results in a  smaller rate. 
The transfer rates for two electrons or holes on an AT-AT dimer are exponentially suppressed  and several orders of magnitude smaller than those on a GC-GC dimer.

At low temperatures  the mean-field decoupling  of the bath modes is not applicable any longer. 
The quantum mechanical nature of bath modes must be taken into account, and
nuclear tunneling is  important: now coherent collective oscillations of the electrons become possible.
As a consequence, the transfer characteristics is changed, and 
the rates can become larger than in the thermal activation regime.

We employ the TD-NRG  to calculate  the population probabilities $d_D(t)$ and $n_{DA}(t)$  
at a temperature $T=1.5 \cdot 10^{-8} eV$  for the parameters of the 
different base pairs coupled to an ohmic bath.  The real-time dynamics for short-time scales is
plotted in  Fig.~\ref{fig_QuantumU}. The transfer shows an incoherent behavior 
with a few remains of oscillations. The decay  rates are  evaluated by fitting the result to the kinetic equations Eq.~(\ref{kinetic}). These fitted rates are added to Fig.~\ref{U4double} as (red) points and are also depicted in Fig.~\ref{fig_rateW}.

The largest rate is found for GC-GC dimer with two electrons, $k_{DA}=0.167\cdot 10^{15} 1/s$, and 
the equilibrium is reached  after about 40 fs,
while the rate for  GC-GC (2h) is  25 times smaller corresponding to a characteristic time scale of $146$fs. 
For $T\to 0$, the optimal regime is already
reached for  $\tilde{U}_{\rm eff}<E_{\alpha 1}$ and, therefore,  earlier than predicted by the Marcus theory.  
Such behavior has also been reported in the biased spin-boson model at low temperatures. \cite{Xu199491} 
Nevertheless, the Marcus rates and the TD-NRG rates  are of the same 
order of magnitude in the optimal regime. Using the classification of the Marcus theory, the transfer in GC-GC (2e) then would be already in the  inverted region at low temperatures.

We would like to emphasize that these transfer rates are significantly larger than those
reported for the single charge subspace. The incoherent relaxation occurs on time scales between a few fs 
and 50fs while the  envelope function of the single charge dynamics  decays on times scales larger than 200fs
as can be seen in Fig.~\ref{fig_single}.

For AT-AT (2e) the two electrons on the donor are self-trapped in an ohmic bath. 
For AT-AT (2h) one hole is transferred with a characteristic time scale of 146fs. 
The transfer is much faster than in the thermal activation regime where the rate is exponentially suppressed.

A summary of the TD-NRG results is included in Fig.~\ref{U4double}. 
We can identify three regions. Below $\tilde{U}_{\rm eff}=0.5eV$ (left of the dotted line)  the system is localized\cite{refId}  for an ohmic bath, and the charges are self-trapped. 
Right of the dotted line -- white region  -- we found  sequential  transfer. 
With increasing $\tilde{U}_{\rm eff}$ more coherent transfer is observed, and the fraction of pair transfer is increasing.

At low temperatures pair transfer would only be present when $\tilde{U}_{\rm eff}$ exceeds $4 eV$  
for the given reorganization energy $E_{\alpha 1} $ in an ohmic bath. 
The charges remain trapped at the donor in the AT-AT (2e) dimer  coupled to an ohmic bath since $\tilde{U}_{\rm eff}=-1eV$  lies in the localized regime. This self-trapping is lifted in a super-ohmic bath: 
pair transfer is observed in  super-ohmic baths as shown in Fig.~\ref{fintepairW}.

\subsection{Finite bond-bond interaction}
\begin{figure}
\vspace*{0cm} \hspace*{0cm} \epsfxsize=9cm
\centerline{\epsffile{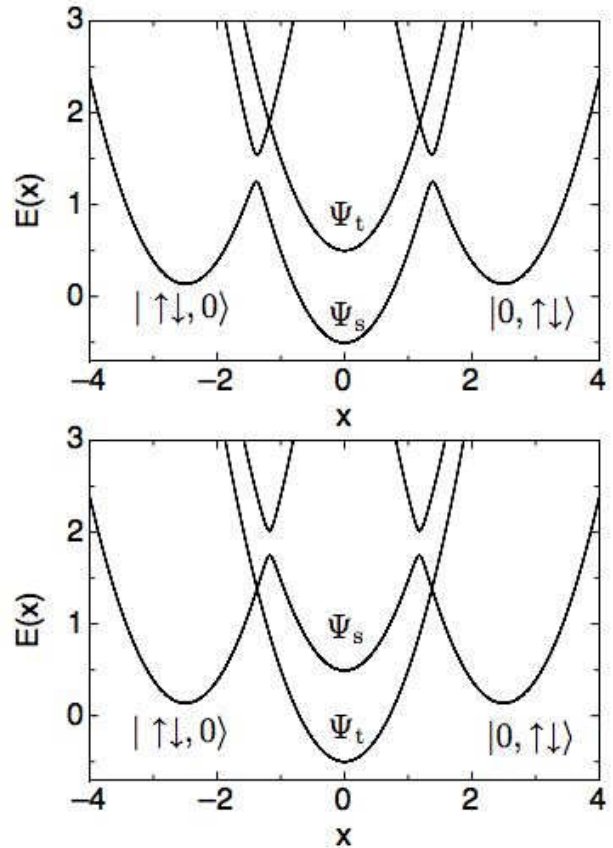}}
\caption{ Schematic potential surface $E(x)$ of the lowest states for a single bosonic mode as a function of the displacement $x$ in arbitrary units for a finite $|W|  \geq\tilde{U}_{\rm eff}$: for $W<0$ (upper panel) and $W>0$ (lower panel). In the latter case $\Psi_t$ is the state with the lowest energy. The hopping matrix element between this state and the initial state $|\uparrow \downarrow, 0\rangle$ is zero.
} \label{fig_marcusW}
\end{figure}
\begin{figure}
\vspace*{0cm} \hspace*{0cm} \epsfxsize=9cm
\centerline{\epsffile{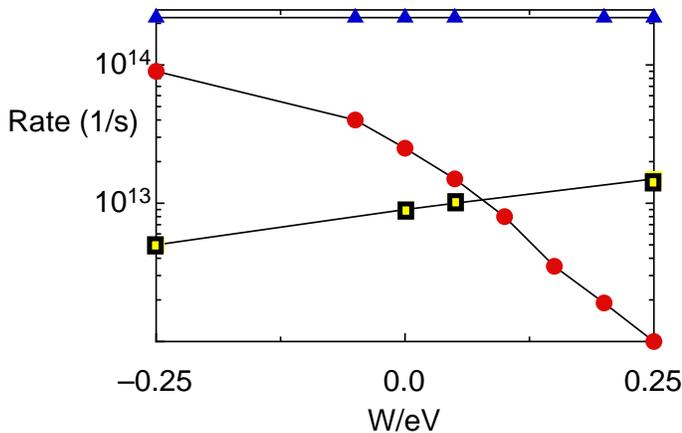}}
\caption{(Color online) 
Low temperature rate $k_{DA}$ that at least one charge is migrating from D to A for  AT-AT(2h) (red circles), GC-GC (2e) (blue triangles) and GC-GC (2 holes) (yellow squares). The time-dependent population probabilities are calculated with the TD-NRG using an ohmic bath and $\alpha=0.5$. As the transfer is in its incoherent regime the rates are obtained by fitting the kinetic equations eq. (\ref{kinetic}). 
} \label{fig_rateW}
\end{figure}

\begin{figure}
\vspace*{0cm} \hspace*{0cm} \epsfxsize=9cm
\centerline{\epsffile{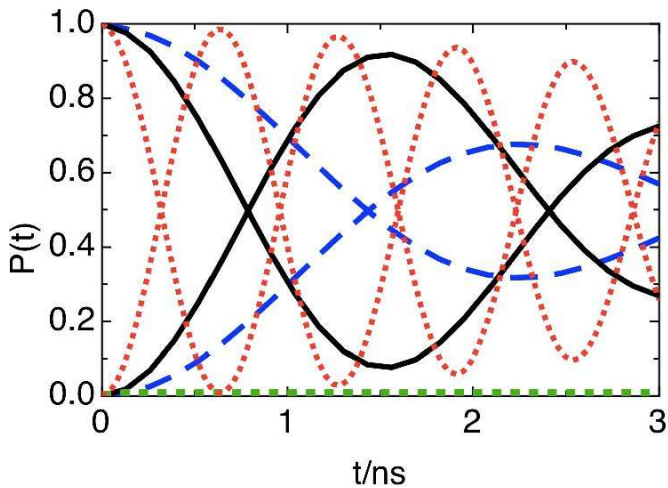}}
\caption{(Color online) 
Low-temperature population probability $P(t)=d_D(t)$ of the doubly occupied donor (starts at one) and acceptor $P(t)=d_{A}(t)$ of the covalent states (starts at zero) for $W=0$ (blue dashed line), 
$W=0.1 eV$ (red dotted line) and $W=-0.1 eV$ (black solid line) as a function of time for AT-AT (2e) considering a super-ohmic bath ($s=2$) and $\alpha=1$. The population probability of the covalent states $P(t)=n_{DA}(t)$ is zero (green squares). 
} \label{fintepairW}
\end{figure}

The bond-bond interaction has two effects: 
First it induces an effective pair hopping between $\ket{\!\uparrow \downarrow,0}$ 
and $\ket{0,\uparrow \downarrow} $  and second it lifts the degeneracy between the triplet and singlet states 
$\ket{\Psi_t}$ and $\ket{\Psi_s}$, respectively. 
The triplet state, which is decoupled from the other states,  becomes  the ground state for a finite $W>0$ and $\tilde{U}_{\rm eff}>0$. Since spin conservation prohibits its population when starting initially 
with two fermions on the donor site, 
the transfer involves only the covalent singlet state $\ket{\Psi_s}$. Its  dynamics is governed by the energy difference $
\Delta E=\tilde{U}_{\rm eff}-W$ between $|\!\uparrow \downarrow,0\rangle$ and $\ket{\Psi_s}$.

For a small $\tilde{U}_{\rm eff}=0.1 eV$ as  in AT-AT(2h) 
the rate predicted by the Marcus theory decreases with increasing $W$ because the energy difference changes its sign from $\tilde{U}_{\rm eff}-W>0$ to $\tilde{U}_{\rm eff}-W<0$. These two regimes are shown schematically 
in Fig. \ref{fig_marcusW} for $|W|>\tilde{U}_{\rm eff}$. 
The dependency of the rate $k_{DA}$ on $W$
obtained  at low temperatures using the TD-NRG follows the prediction of the Marcus theory as shown
in Fig.~\ref{fig_rateW}  as (red) circles.

As discussed above, the  GC-GC (2h) dimer is located in the  inverted
region as a
consequence 
of the large $\tilde{U}_{\rm eff}=2 eV$. 
An increasing $W$ reduces the energy difference between the initial state and $\ket{\Psi_s}$ leading to an increase of $k_{DA}$ as depicted in Fig.~\ref{fig_rateW} (squares).
The transfer in GC-GC (2e) shows almost no dependence on $W$ (Fig.~\ref{fig_rateW}, triangles) since it is located in the activationless regime.

An electron-pair in an AT-AT dimer remains  self-trapped for  $|W|<0.25 eV$ in an ohmic bath.
In a super-ohmic bath ($s>1$), however, coherent transfer is found at low temperatures.
The electrons oscillate coherently as pairs  between the donor and the acceptor while the high lying 
singlet covalent state $\ket{\Psi_s}$  is only virtually occupied. The oscillation  frequency is approximately
given by $\frac{4\Delta_{\rm eff}^2}{|\tilde{U}_{\rm eff}-W|}+2 W_{\rm eff}$ where $W_{\rm eff}$ is the renormalized pair-hopping in the presence of the bosonic bath:
this frequency increases with increasing $W$. 

The occupation probabilities 
show slow oscillations with a period  of the order of $1-4$ ns as depicted in Fig.~\ref{fintepairW}.
Note that the time scale is multiple orders of magnitude larger
compared to 
the fs time-scales
of the GC-GC dimer.
The relaxation rate of the envelope function corresponds to a significantly longer time scale $\approx 50ns$.

\section{Summary}
\label{sum}

We have calculated the real-time dynamics of a redox system
with multiple charges.
At  temperatures lower than the activation energy we have used 
the time-dependent numerical renormalization group approach; 
at temperatures larger than the activation energy when the transfer is
governed by thermal activation we have employed  
the Marcus theory. 
Four different DNA dimers have been considered: GC-GC and  AT-AT with initially two excess electrons or holes on 
the donor site. All results are obtained for realistic matrix elements obtained by ab-initio methods.\cite{Starikov}

The charge transfer characteristics  of a system with two excess charges deviates substantially from
those of a single excess charge.  The transfer  occurs correlated and 
depends strongly on diagonal and off-diagonal Coulomb matrix
elements. We predict that an additional charge in DNA sensors or DNA devices
will show base-pair specific signal amplification or attenuation.

We have calculated the rate $k_{DA}$ that at least one charge is transferred 
from the donor to the acceptor. 
The time-dependent population probabilities of the singlet covalent state ($\ket{\Psi_s}$) 
and two ionic states ($|\!\uparrow \downarrow,0 \rangle$, $|0,\uparrow \downarrow \rangle$)
depend on the difference between the effective diagonal Coulomb
interaction $\tilde{U}_{\rm eff}$ and  the off-diagonal bond-bond interaction $W$.

We have found a qualitative agreement between the Marcus theory a high-temperatures and the TD-NRG at 
low temperatures. The transfer rate $k_{DA}$ shows a dome-shape as
function 
of $\tilde{U}_{\rm eff}$ in the Marcus theory as
depicted in Fig.~\ref{U4double}. It is slightly shifted to 
lower values of $\tilde{U}_{\rm eff}$ and broadened at low temperatures
as calculated via TD-NRG. However, self-trapping is only found in the TD-NRG.
The correct
application of the Marcus theory requires the restriction to transitions compatible with
spin-conservation: the triplet state $\ket{\Psi_t}$ remains inaccessible since we always start with two charges
in a spin-singlet state on the donor.


The maximal transfer rate $k_{DA}$, defining the activation-less regime, is significantly larger in the subspace of two excess charges 
compared to the subspace of a single excess charge. At low temperatures, the largest rate is found for
two excess electrons in GC-GC and for two excess-holes at room temperature. 
Electron-pairs initially on the donor in an AT-AT dimer are either self-trapped (ohmic bath) or are slowly 
oscillating as pairs between D and A (super-ohmic bath) at low temperatures.

The dependency of the transfer rate on a finite but small bond-bond interaction $W$ 
can be understood in terms of a simple shift of $U_{\rm eff} \to U_{\rm eff}-W$: The 
rate $k_{DA}$ now peaks at $\approx U_{\rm eff}-W-E_{\alpha 1}$.
Consequently, the transfer rate increases for
two holes on GC-GC which is located in the inverted regime while $k_{DA}$ decreases for
two holes on AT-AT with increasing $W$. It remains constant for two electrons  on GC-GC since the rate
appears to be rather insensitive to $W$ in the activation-less regime.

Furthermore, the presence of the additional 
bond-bond interaction $W$  splits the degeneracy of the singlet and the triplet state,
$\ket{\Psi_s}$ and $\ket{\Psi_t}$, respectively. Without any further spin relaxation process, the system cannot
reach the electronic thermodynamic ground state  $\ket{\Psi_t}$ for $W>0$ and $U_{\rm eff}>0$ since the triplet state remains
unpopulated at all times. The system remains trapped in an intermediate metastable state whose life-time will
depend on the spin relaxation mechanism not considered here.

The transfer characteristics of two charges in a DNA dimer depends strongly on the base pair 
specific Coulomb interactions.  These effects will become important  when such 
redox systems are weakly contacted. They strongly influence the current-voltage characteristics 
once multiple excess charges are relevant at  higher  source-drain voltages.\cite{TornowGoldenrule}

\section*{Acknowledgment}
This research was supported by the State of Lower- 
Saxony and the Volkswagen Foundation (ST,GZ) and
by the DFG through
SFB 608 (RB) and An275/6-2 (FBA). We acknowledge helpful discussions with E. Starikow, A. Nitzan and A. Schiller.


\begin{thebibliography}{40}
\expandafter\ifx\csname natexlab\endcsname\relax\def\natexlab#1{#1}\fi
\expandafter\ifx\csname bibnamefont\endcsname\relax
  \def\bibnamefont#1{#1}\fi
\expandafter\ifx\csname bibfnamefont\endcsname\relax
  \def\bibfnamefont#1{#1}\fi
\expandafter\ifx\csname citenamefont\endcsname\relax
  \def\citenamefont#1{#1}\fi
\expandafter\ifx\csname url\endcsname\relax
  \def\url#1{\texttt{#1}}\fi
\expandafter\ifx\csname urlprefix\endcsname\relax\def\urlprefix{URL }\fi
\providecommand{\bibinfo}[2]{#2}
\providecommand{\eprint}[2][]{\url{#2}}

\bibitem[{\citenamefont{Chakraboty}(2007)}]{Chakra}
\bibinfo{author}{\bibfnamefont{T.}~\bibnamefont{Chakraboty}},
  \emph{\bibinfo{title}{Charge Migration in DNA: Perspectives from Physics,
  Chemistry and Charge Migration in DNA: Perspectives from Physics, Chemistry
  and Biology}} (\bibinfo{publisher}{Springer}, \bibinfo{year}{2007}).

\bibitem[{\citenamefont{Schuster}(2004)}]{Schuster}
\bibinfo{author}{\bibfnamefont{G.}~\bibnamefont{Schuster}},
  \emph{\bibinfo{title}{Long-Range Charge Transfer in DNA}}
  (\bibinfo{publisher}{Springer}, \bibinfo{year}{2004}).

\bibitem[{\citenamefont{Wagenknecht}(2005)}]{Wagenknecht}
\bibinfo{author}{\bibnamefont{Wagenknecht}}, \emph{\bibinfo{title}{Charge
  Transfer in DNA. From Mechanism to Application}}
  (\bibinfo{publisher}{Wiley-VCH}, \bibinfo{year}{2005}).

\bibitem[{\citenamefont{Starikov}(2006)}]{Starikov2}
\bibinfo{author}{\bibfnamefont{E.}~\bibnamefont{Starikov}},
  \emph{\bibinfo{title}{Modern Methods for Theoretical Physical Chemistry of
  Biopolymers}} (\bibinfo{publisher}{Elsevier Science}, \bibinfo{year}{2006}).

\bibitem[{\citenamefont{Endres et~al.}(2004)\citenamefont{Endres, Cox, and
  Singh}}]{Endres:2004vg}
\bibinfo{author}{\bibfnamefont{R.~G.} \bibnamefont{Endres}},
  \bibinfo{author}{\bibfnamefont{D.~L.} \bibnamefont{Cox}}, \bibnamefont{and}
  \bibinfo{author}{\bibfnamefont{R.~R.~P.} \bibnamefont{Singh}},
  \bibinfo{journal}{Rev. Mod. Phys.} \textbf{\bibinfo{volume}{76}},
  \bibinfo{pages}{195} (\bibinfo{year}{2004}).

\bibitem[{\citenamefont{Lu et~al.}(2007)\citenamefont{Lu, Kalantari, and
  Wang}}]{platin}
\bibinfo{author}{\bibfnamefont{Q.-B.} \bibnamefont{Lu}},
  \bibinfo{author}{\bibfnamefont{S.}~\bibnamefont{Kalantari}},
  \bibnamefont{and} \bibinfo{author}{\bibfnamefont{C.-R.} \bibnamefont{Wang}},
  \bibinfo{journal}{Molecular Pharmaceutics} \textbf{\bibinfo{volume}{4}},
  \bibinfo{pages}{624} (\bibinfo{year}{2007}).

\bibitem[{\citenamefont{Giese}(2002)}]{Giese:2002le}
\bibinfo{author}{\bibfnamefont{B.}~\bibnamefont{Giese}},
  \bibinfo{journal}{Annual Review of Biochemistry}
  \textbf{\bibinfo{volume}{71}}, \bibinfo{pages}{51} (\bibinfo{year}{2002}).

\bibitem[{\citenamefont{Delaney and Barton}(2003)}]{Delaney:2003wo}
\bibinfo{author}{\bibfnamefont{S.}~\bibnamefont{Delaney}} \bibnamefont{and}
  \bibinfo{author}{\bibfnamefont{J.~K.} \bibnamefont{Barton}},
  \bibinfo{journal}{The Journal of Organic Chemistry}
  \textbf{\bibinfo{volume}{68}}, \bibinfo{pages}{6475} (\bibinfo{year}{2003}).

\bibitem[{\citenamefont{Gutierrez et~al.}(2009)\citenamefont{Gutierrez,
  Caetano, Woiczikowski, Kubar, Elstner, and Cuniberti}}]{gutierrez:208102}
\bibinfo{author}{\bibfnamefont{R.}~\bibnamefont{Gutierrez}},
  \bibinfo{author}{\bibfnamefont{R.~A.} \bibnamefont{Caetano}},
  \bibinfo{author}{\bibfnamefont{B.~P.} \bibnamefont{Woiczikowski}},
  \bibinfo{author}{\bibfnamefont{T.}~\bibnamefont{Kubar}},
  \bibinfo{author}{\bibfnamefont{M.}~\bibnamefont{Elstner}}, \bibnamefont{and}
  \bibinfo{author}{\bibfnamefont{G.}~\bibnamefont{Cuniberti}},
  \bibinfo{journal}{Physical Review Letters} \textbf{\bibinfo{volume}{102}},
  \bibinfo{eid}{208102} (pages~\bibinfo{numpages}{4}) (\bibinfo{year}{2009}).

\bibitem[{\citenamefont{Bulla et~al.}(2006)\citenamefont{Bulla, Guti\'errez,
  and Cuniberti}}]{Bulla:2006tx}
\bibinfo{author}{\bibfnamefont{R.}~\bibnamefont{Bulla}},
  \bibinfo{author}{\bibfnamefont{R.}~\bibnamefont{Guti\'errez}},
  \bibnamefont{and}
  \bibinfo{author}{\bibfnamefont{G.}~\bibnamefont{Cuniberti}}, in
  \emph{\bibinfo{booktitle}{Modern methods for theoretical physical chemistry
  of biopolymers}}, edited by
  \bibinfo{editor}{\bibfnamefont{E.}~\bibnamefont{Starikov}},
  \bibinfo{editor}{\bibfnamefont{J.~P.} \bibnamefont{Lewis}}, \bibnamefont{and}
  \bibinfo{editor}{\bibfnamefont{S.}~\bibnamefont{Tanaka}}
  (\bibinfo{publisher}{Elsevier}, \bibinfo{address}{Amsterdam},
  \bibinfo{year}{2006}), pp. \bibinfo{pages}{379--391}.

\bibitem[{\citenamefont{Marcus}(1956)}]{marcus:966}
\bibinfo{author}{\bibfnamefont{R.~A.} \bibnamefont{Marcus}},
  \bibinfo{journal}{The Journal of Chemical Physics}
  \textbf{\bibinfo{volume}{24}}, \bibinfo{pages}{966} (\bibinfo{year}{1956}).

\bibitem[{\citenamefont{DeVault}(1984)}]{DeVault:1984rg}
\bibinfo{author}{\bibfnamefont{D.}~\bibnamefont{DeVault}},
  \emph{\bibinfo{title}{Quantum-mechanical tunnelling in biological systems}}
  (\bibinfo{publisher}{Cambridge University Press, Cambridge},
  \bibinfo{year}{1984}).

\bibitem[{\citenamefont{Barone et~al.}(1992)\citenamefont{Barone, Smith, and
  Galvo}}]{Barone:1992sw}
\bibinfo{author}{\bibfnamefont{P.~M. V.~B.} \bibnamefont{Barone}},
  \bibinfo{author}{\bibfnamefont{C.~M.} \bibnamefont{Smith}}, \bibnamefont{and}
  \bibinfo{author}{\bibfnamefont{D.~S.} \bibnamefont{Galvo}},
  \bibinfo{journal}{Phys. Rev. A} \textbf{\bibinfo{volume}{45}},
  \bibinfo{pages}{3592} (\bibinfo{year}{1992}).

\bibitem[{\citenamefont{Leggett et~al.}(1987)\citenamefont{Leggett,
  Chakravarty, Dorsey, Fisher, Garg, and Zwerger}}]{Leggett:1987gc}
\bibinfo{author}{\bibfnamefont{A.~J.} \bibnamefont{Leggett}},
  \bibinfo{author}{\bibfnamefont{S.}~\bibnamefont{Chakravarty}},
  \bibinfo{author}{\bibfnamefont{A.~T.} \bibnamefont{Dorsey}},
  \bibinfo{author}{\bibfnamefont{M.~P.~A.} \bibnamefont{Fisher}},
  \bibinfo{author}{\bibfnamefont{A.}~\bibnamefont{Garg}}, \bibnamefont{and}
  \bibinfo{author}{\bibfnamefont{W.}~\bibnamefont{Zwerger}},
  \bibinfo{journal}{Rev. Mod. Phys.} \textbf{\bibinfo{volume}{59}},
  \bibinfo{pages}{1} (\bibinfo{year}{1987}).

\bibitem[{\citenamefont{Evans}(2008)}]{Evans:2008ij}
\bibinfo{author}{\bibfnamefont{D.~H.} \bibnamefont{Evans}},
  \bibinfo{journal}{Chemical Reviews} \textbf{\bibinfo{volume}{108}},
  \bibinfo{pages}{2113} (\bibinfo{year}{2008}).

\bibitem[{\citenamefont{M\"uhlbacher et~al.}(2005)\citenamefont{M\"uhlbacher,
  Ankerhold, and Komnik}}]{Muhlbacher:2005gn}
\bibinfo{author}{\bibfnamefont{L.}~\bibnamefont{M\"uhlbacher}},
  \bibinfo{author}{\bibfnamefont{J.}~\bibnamefont{Ankerhold}},
  \bibnamefont{and} \bibinfo{author}{\bibfnamefont{A.}~\bibnamefont{Komnik}},
  \bibinfo{journal}{Phys. Rev. Lett.} \textbf{\bibinfo{volume}{95}},
  \bibinfo{pages}{220404} (\bibinfo{year}{2005}).

\bibitem[{\citenamefont{Tornow et~al.}(2008)\citenamefont{Tornow, Bulla,
  Anders, and Nitzan}}]{tornow:035434}
\bibinfo{author}{\bibfnamefont{S.}~\bibnamefont{Tornow}},
  \bibinfo{author}{\bibfnamefont{R.}~\bibnamefont{Bulla}},
  \bibinfo{author}{\bibfnamefont{F.~B.} \bibnamefont{Anders}},
  \bibnamefont{and} \bibinfo{author}{\bibfnamefont{A.}~\bibnamefont{Nitzan}},
  \bibinfo{journal}{Physical Review B (Condensed Matter and Materials Physics)}
  \textbf{\bibinfo{volume}{78}}, \bibinfo{eid}{035434}
  (pages~\bibinfo{numpages}{14}) (\bibinfo{year}{2008}).

\bibitem[{\citenamefont{Zusman and Beratan}(1996)}]{zusman:165}
\bibinfo{author}{\bibfnamefont{L.~D.} \bibnamefont{Zusman}} \bibnamefont{and}
  \bibinfo{author}{\bibfnamefont{D.~N.} \bibnamefont{Beratan}},
  \bibinfo{journal}{The Journal of Chemical Physics}
  \textbf{\bibinfo{volume}{105}}, \bibinfo{pages}{165} (\bibinfo{year}{1996}).

\bibitem[{\citenamefont{Zusman and Beratan}(1997)}]{Zusman:1997qe}
\bibinfo{author}{\bibfnamefont{L.~D.} \bibnamefont{Zusman}} \bibnamefont{and}
  \bibinfo{author}{\bibfnamefont{D.~N.} \bibnamefont{Beratan}},
  \bibinfo{journal}{The Journal of Physical Chemistry A}
  \textbf{\bibinfo{volume}{101}}, \bibinfo{pages}{4136} (\bibinfo{year}{1997}).

\bibitem[{\citenamefont{Petrov et~al.}(2003)\citenamefont{Petrov, Teslenko, and
  May}}]{Petrov:2003qt}
\bibinfo{author}{\bibfnamefont{E.~G.} \bibnamefont{Petrov}},
  \bibinfo{author}{\bibfnamefont{V.~I.} \bibnamefont{Teslenko}},
  \bibnamefont{and} \bibinfo{author}{\bibfnamefont{V.}~\bibnamefont{May}},
  \bibinfo{journal}{Phys. Rev. E} \textbf{\bibinfo{volume}{68}},
  \bibinfo{pages}{061916} (\bibinfo{year}{2003}).

\bibitem[{\citenamefont{Tornow et~al.}(2006)\citenamefont{Tornow, Tong, and
  Bulla}}]{refId}
\bibinfo{author}{\bibfnamefont{S.}~\bibnamefont{Tornow}},
  \bibinfo{author}{\bibfnamefont{N.-H.} \bibnamefont{Tong}}, \bibnamefont{and}
  \bibinfo{author}{\bibfnamefont{R.}~\bibnamefont{Bulla}},
  \bibinfo{journal}{Europhysics Letters} \textbf{\bibinfo{volume}{73}},
  \bibinfo{pages}{913} (\bibinfo{year}{2006}).

\bibitem[{\citenamefont{Starikov}(20031)}]{Starikov}
\bibinfo{author}{\bibfnamefont{E.~B.} \bibnamefont{Starikov}},
  \bibinfo{journal}{Philosophical Magazine Letters}
  \textbf{\bibinfo{volume}{83}}, \bibinfo{pages}{699} (\bibinfo{year}{20031}).

\bibitem[{\citenamefont{Xu et~al.}(2004)\citenamefont{Xu, Zhang, Li, and
  Tao}}]{Xu:2004wq}
\bibinfo{author}{\bibnamefont{Xu}}, \bibinfo{author}{\bibnamefont{Zhang}},
  \bibinfo{author}{\bibnamefont{Li}}, \bibnamefont{and}
  \bibinfo{author}{\bibnamefont{Tao}}, \bibinfo{journal}{Nano Letters}
  \textbf{\bibinfo{volume}{4}}, \bibinfo{pages}{1105} (\bibinfo{year}{2004}).

\bibitem[{\citenamefont{Nitzan}(2001)}]{Nitzan:2001ws}
\bibinfo{author}{\bibfnamefont{A.}~\bibnamefont{Nitzan}}, \bibinfo{journal}{The
  Journal of Physical Chemistry A} \textbf{\bibinfo{volume}{105}},
  \bibinfo{pages}{2677} (\bibinfo{year}{2001}).

\bibitem[{\citenamefont{Tornow and Zwicknagl}(2010)}]{TornowGoldenrule}
\bibinfo{author}{\bibfnamefont{S.}~\bibnamefont{Tornow}} \bibnamefont{and}
  \bibinfo{author}{\bibfnamefont{G.}~\bibnamefont{Zwicknagl}},
  \bibinfo{journal}{physica status solidi (RRL) - Rapid Research Letters}
  \textbf{\bibinfo{volume}{4}}, \bibinfo{pages}{46} (\bibinfo{year}{2010}).

\bibitem[{\citenamefont{Pratviel and
  Meunier}(2006)}]{GeneviEve-Pratviel:2006kn}
\bibinfo{author}{\bibfnamefont{G.}~\bibnamefont{Pratviel}} \bibnamefont{and}
  \bibinfo{author}{\bibfnamefont{B.}~\bibnamefont{Meunier}},
  \bibinfo{journal}{Chemistry - A European Journal}
  \textbf{\bibinfo{volume}{12}}, \bibinfo{pages}{6018} (\bibinfo{year}{2006}).

\bibitem[{\citenamefont{Fan et~al.}(2005)\citenamefont{Fan, Plaxco, and
  Heeger}}]{Fan2005186}
\bibinfo{author}{\bibfnamefont{C.}~\bibnamefont{Fan}},
  \bibinfo{author}{\bibfnamefont{K.~W.} \bibnamefont{Plaxco}},
  \bibnamefont{and} \bibinfo{author}{\bibfnamefont{A.~J.}
  \bibnamefont{Heeger}}, \bibinfo{journal}{Trends in Biotechnology}
  \textbf{\bibinfo{volume}{23}}, \bibinfo{pages}{186 } (\bibinfo{year}{2005}).

\bibitem[{\citenamefont{Anders and Schiller}(2006)}]{anders:245113}
\bibinfo{author}{\bibfnamefont{F.~B.} \bibnamefont{Anders}} \bibnamefont{and}
  \bibinfo{author}{\bibfnamefont{A.}~\bibnamefont{Schiller}},
  \bibinfo{journal}{Physical Review B (Condensed Matter and Materials Physics)}
  \textbf{\bibinfo{volume}{74}}, \bibinfo{eid}{245113}
  (pages~\bibinfo{numpages}{22}) (\bibinfo{year}{2006}).

\bibitem[{\citenamefont{Anders and Schiller}(2005)}]{Anders:2005lm}
\bibinfo{author}{\bibfnamefont{F.~B.} \bibnamefont{Anders}} \bibnamefont{and}
  \bibinfo{author}{\bibfnamefont{A.}~\bibnamefont{Schiller}},
  \bibinfo{journal}{Phys. Rev. Lett.} \textbf{\bibinfo{volume}{95}},
  \bibinfo{pages}{196801} (\bibinfo{year}{2005}).

\bibitem[{\citenamefont{Krishna-murthy
  et~al.}(1980)\citenamefont{Krishna-murthy, Wilkins, and
  Wilson}}]{Krishna-murthy:1980bz}
\bibinfo{author}{\bibfnamefont{H.~R.} \bibnamefont{Krishna-murthy}},
  \bibinfo{author}{\bibfnamefont{J.~W.} \bibnamefont{Wilkins}},
  \bibnamefont{and} \bibinfo{author}{\bibfnamefont{K.~G.}
  \bibnamefont{Wilson}}, \bibinfo{journal}{Phys. Rev. B}
  \textbf{\bibinfo{volume}{21}}, \bibinfo{pages}{1003} (\bibinfo{year}{1980}).

\bibitem[{\citenamefont{Bulla et~al.}(2005)\citenamefont{Bulla, Lee, Tong, and
  Vojta}}]{bulla:045122}
\bibinfo{author}{\bibfnamefont{R.}~\bibnamefont{Bulla}},
  \bibinfo{author}{\bibfnamefont{H.-J.} \bibnamefont{Lee}},
  \bibinfo{author}{\bibfnamefont{N.-H.} \bibnamefont{Tong}}, \bibnamefont{and}
  \bibinfo{author}{\bibfnamefont{M.}~\bibnamefont{Vojta}},
  \bibinfo{journal}{Physical Review B (Condensed Matter and Materials Physics)}
  \textbf{\bibinfo{volume}{71}}, \bibinfo{eid}{045122} (\bibinfo{year}{2005}).

\bibitem[{\citenamefont{Bulla et~al.}(2003)\citenamefont{Bulla, Tong, and
  Vojta}}]{Bulla:2003pw}
\bibinfo{author}{\bibfnamefont{R.}~\bibnamefont{Bulla}},
  \bibinfo{author}{\bibfnamefont{N.-H.} \bibnamefont{Tong}}, \bibnamefont{and}
  \bibinfo{author}{\bibfnamefont{M.}~\bibnamefont{Vojta}},
  \bibinfo{journal}{Phys. Rev. Lett.} \textbf{\bibinfo{volume}{91}},
  \bibinfo{pages}{170601} (\bibinfo{year}{2003}).

\bibitem[{\citenamefont{Bulla et~al.}(2008)\citenamefont{Bulla, Costi, and
  Pruschke}}]{bulla:395}
\bibinfo{author}{\bibfnamefont{R.}~\bibnamefont{Bulla}},
  \bibinfo{author}{\bibfnamefont{T.~A.} \bibnamefont{Costi}}, \bibnamefont{and}
  \bibinfo{author}{\bibfnamefont{T.}~\bibnamefont{Pruschke}},
  \bibinfo{journal}{Reviews of Modern Physics} \textbf{\bibinfo{volume}{80}},
  \bibinfo{eid}{395} (pages~\bibinfo{numpages}{56}) (\bibinfo{year}{2008}).

\bibitem[{\citenamefont{Nitzan}(2006)}]{Nitzan_book}
\bibinfo{author}{\bibfnamefont{A.}~\bibnamefont{Nitzan}},
  \emph{\bibinfo{title}{Chemical Dynamics in Condensed Phases: Relaxation,
  Transfer, and Reactions in Condensed Molecular System}}
  (\bibinfo{publisher}{Oxford University Press, USA}, \bibinfo{year}{2006}).

\bibitem[{\citenamefont{Garg et~al.}(1985)\citenamefont{Garg, Onuchic, and
  Ambegaokar}}]{garg:4491}
\bibinfo{author}{\bibfnamefont{A.}~\bibnamefont{Garg}},
  \bibinfo{author}{\bibfnamefont{J.~N.} \bibnamefont{Onuchic}},
  \bibnamefont{and}
  \bibinfo{author}{\bibfnamefont{V.}~\bibnamefont{Ambegaokar}},
  \bibinfo{journal}{The Journal of Chemical Physics}
  \textbf{\bibinfo{volume}{83}}, \bibinfo{pages}{4491} (\bibinfo{year}{1985}).

\bibitem[{\citenamefont{Weiss}(1999)}]{Weiss_book}
\bibinfo{author}{\bibfnamefont{U.}~\bibnamefont{Weiss}},
  \emph{\bibinfo{title}{Quantum dissipative systems}}
  (\bibinfo{publisher}{World Scientific, Singapore}, \bibinfo{year}{1999}).

\bibitem[{\citenamefont{Vladimirov et~al.}(2008)\citenamefont{Vladimirov,
  Ivanova, and Rosch}}]{vladimirov:194515}
\bibinfo{author}{\bibfnamefont{E.}~\bibnamefont{Vladimirov}},
  \bibinfo{author}{\bibfnamefont{A.}~\bibnamefont{Ivanova}}, \bibnamefont{and}
  \bibinfo{author}{\bibfnamefont{N.}~\bibnamefont{Rosch}},
  \bibinfo{journal}{The Journal of Chemical Physics}
  \textbf{\bibinfo{volume}{129}}, \bibinfo{eid}{194515} (\bibinfo{year}{2008}).

\bibitem[{\citenamefont{Kleinekath{\"o}fer
  et~al.}(2006)\citenamefont{Kleinekath{\"o}fer, Li, and Schreiber}}]{Kleink}
\bibinfo{author}{\bibfnamefont{U.}~\bibnamefont{Kleinekath{\"o}fer}},
  \bibinfo{author}{\bibfnamefont{G.}~\bibnamefont{Li}}, \bibnamefont{and}
  \bibinfo{author}{\bibfnamefont{M.}~\bibnamefont{Schreiber}},
  \bibinfo{journal}{Journal of Luminescence} \textbf{\bibinfo{volume}{119-120}}
  (\bibinfo{year}{2006}).

\bibitem[{\citenamefont{Praprotnik et~al.}(2004)\citenamefont{Praprotnik,
  Janezi, and Mavri}}]{Praprotnik:2004yf}
\bibinfo{author}{\bibfnamefont{M.}~\bibnamefont{Praprotnik}},
  \bibinfo{author}{\bibfnamefont{D.}~\bibnamefont{Janezi}}, \bibnamefont{and}
  \bibinfo{author}{\bibfnamefont{J.}~\bibnamefont{Mavri}},
  \bibinfo{journal}{The Journal of Physical Chemistry A}
  \textbf{\bibinfo{volume}{108}}, \bibinfo{pages}{11056}
  (\bibinfo{year}{2004}).

\bibitem[{\citenamefont{Xu and Schulten}(1994)}]{Xu199491}
\bibinfo{author}{\bibfnamefont{D.}~\bibnamefont{Xu}} \bibnamefont{and}
  \bibinfo{author}{\bibfnamefont{K.}~\bibnamefont{Schulten}},
  \bibinfo{journal}{Chemical Physics} \textbf{\bibinfo{volume}{182}},
  \bibinfo{pages}{91 } (\bibinfo{year}{1994}).

\end{thebibliography}
\end{document}